\documentclass[sigconf]{acmart} 

\AtBeginDocument{%
  }

\copyrightyear{2024}
\acmYear{2024}
\setcopyright{acmlicensed}\acmConference[CHASE '24]{2024 IEEE/ACM 17th International Conference on Cooperative and Human Aspects of Software Engineering }{April 14--15, 2024}{Lisbon, Portugal}
\acmBooktitle{2024 IEEE/ACM 17th International Conference on Cooperative and Human Aspects of Software Engineering (CHASE '24), April 14--15, 2024, Lisbon, Portugal}
\acmDOI{10.1145/3641822.3641869}
\acmISBN{979-8-4007-0533-5/24/04}

\usepackage{subcaption}

\usepackage{todonotes}
\setuptodonotes{inline}

\usepackage[framemethod=tikz]{mdframed}
\newmdenv[
  innerleftmargin=7pt,
  innerrightmargin=7pt,
  tikzsetting={draw=black,dashed,line width=0.5pt,dash pattern = on 4pt off 2pt},
  linecolor=white,
  backgroundcolor=white
]{dashedbox}
\newmdenv[
  innerleftmargin=7pt,
  innerrightmargin=7pt,
  tikzsetting={draw=black, line width=0.5pt},
  linecolor=black,
  backgroundcolor=white
]{normalbox}
\newmdenv[
  roundcorner=5pt,
  innerleftmargin=7pt,
  innerrightmargin=7pt,
  tikzsetting={draw=black, line width=0.5pt},
  linecolor=black,
  backgroundcolor=white,
  skipabove=5pt,
  skipbelow=5pt
]{roundedbox}

\newmdenv[
  linewidth=2pt,
  roundcorner=5pt,
  innertopmargin=0pt,
  innerbottommargin=0pt,
]{myframe}

\begin{document}

\title{UX Debt: Developers Borrow While Users Pay}

\author{Sebastian Baltes}
\authornote{This research was initiated while the first author was working at QAware GmbH.}
\email{sebastian.baltes@uni-bayreuth.de}
\orcid{0000-0002-2442-7522}
\affiliation{%
  \institution{University of Bayreuth}
  \country{Germany}
}

\author{Veronika Dashuber}
\email{veronika.dashuber@qaware.de}
\orcid{0000-0001-8577-5646}
\affiliation{%
  \institution{QAware GmbH}
  \country{Germany}
}

\renewcommand{\shortauthors}{Baltes and Dashuber}

\begin{abstract}
Technical debt has become a well-known metaphor among software professionals, illustrating how shortcuts taken during development can accumulate and become a burden for software projects. In the traditional notion of technical debt, software developers borrow from the maintainability and extensibility of a software system for a short-term speed up in development time.
In the future, they are the ones who pay the interest in form of longer development times.
User experience (UX) debt, on the other hand, focuses on shortcuts taken to speed up development at the expense of subpar usability, thus mainly borrowing from user efficiency.
Most research considers code-centric technical debt, focusing on the implementation. 
With this article, we want to build awareness for the often overlooked UX debt of software systems, shifting the focus from the source code towards users.
We outline three classes of UX debt that we observed in practice: code-centric, architecture-centric, and process-centric UX debt.
In an expert survey, we validated those classes, with code-centric and process-centric UX debt getting the strongest support. 
We discuss our participants' feedback and present recommendations on how software development teams can mitigate UX debt in their user-facing applications.
\end{abstract}

\maketitle

\section{Introduction}

Going into \emph{debt} means that a borrower trades additional costs in the form of interest for immediacy, that is, being able to spend money now that they otherwise would not have available.
However, the borrower must be able to pay off debt plus interest in addition to the other future investments they intend to make.
Otherwise, the accumulation of debt and interest leads to bankruptcy.
The term \emph{technical debt}, coined by Ward Cunningham in 1992 \cite{DBLP:journals/oopsm/Cunningham93}, transfers this financial concept to software engineering, describing how shortcuts taken during the development of a software system accumulate and steer more and more development resources towards maintenance.
Taking shortcuts allows teams to ship features earlier and refactor later based on the experience gained during the feature's usage.
The feature that is shipped earlier is an investment that would not be possible without going into debt.
However, the team pays interest in terms of reduced maintainability and extensibility until a refactoring is done and the debt is paid back.

\emph{User experience (UX)} is ``a person's perceptions and responses that result from the use or anticipated use of a product, system or service''~\cite{iso9241-210:2019}.
While Cunningham’s technical debt metaphor focuses on developers who implement and maintain a software system, \emph{UX debt} focuses on the users of the software. The term was introduced by Andrew J Wright in a blog post published in 2013~\cite{Wright2013}, where he defined UX debt as ``the quality gap between the experience your digital product delivers now and the improved experience it could offer given the necessary time and resources.''
Wright distinguishes between \emph{intentional UX debt}, resulting from ``project constraints and deliberate corner-cutting'' and \emph{unintentional UX debt}, resulting from ``misconceptions about users' needs.'' 

In some situations, UX debt can be the result of traditional technical debt, but it can also exist independently of it.
In general, UX debt shifts the focus from the people who maintain a software system to its users.
There are cases where UX debt directly corresponds to traditional technical debt, as debt at the implementation level can ``leak'' into the user-facing parts of an application.
We refer to this direct impact of technical debt on usability as \emph{code-centric UX debt} (Section~\ref{sec:code-centric}).
However, UX debt can exist without the presence of traditional technical debt. This happens, for example, when architectural decisions cause a suboptimal user interface composition, that is, a ``clean'' software architecture causes usability issues in the front-end.
We refer to this class as \emph{architecture-centric UX debt} (Section~\ref{sec:architecture-centric}).
We also discuss \emph{process-centric UX debt}, which takes a more holistic view of the application and captures cases where the software is not aligned with typical user workflows, but again might be completely clean from a technical point of view (Section~\ref{sec:process-centric}).

The three above-mentioned UX debt classes can be, according to Wright's terminology, intentional or unintentional.
An example of \emph{intentional process-centric UX debt} is trading new features for improvements to cumbersome user workflows because they already ``work.'' Such prioritizations are a common part of software development. However, if the choices jeopardize the user experience, they result in UX debt because the users have to continuously follow inefficient workflows, and in that pay the debt for developers' decisions~\cite{Kalbach2014}.
There are situations in which developers deliberately decide against a solution resulting in a better UX.
In other cases, UX debt might be introduced unintentionally.
This happens when developers lack understanding of users' needs and abilities, thus implementing features in a way that is not aligned with users' desired or most efficient workflows.

\begin{figure}
    \centering
    \includegraphics[width=\columnwidth,trim=0pt 0pt 0pt 0pt,clip]{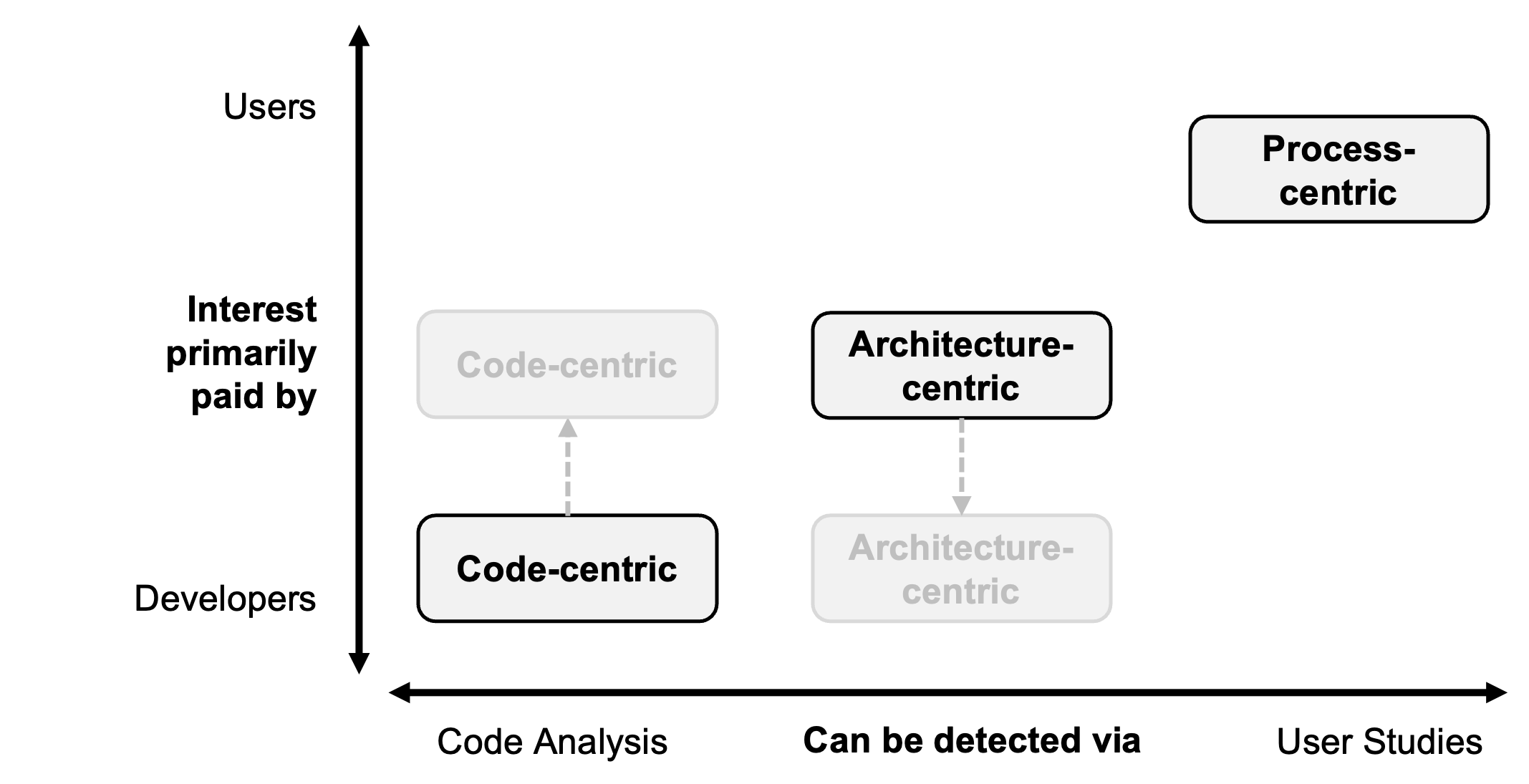}
    \caption{Initial framework of UX debt classes and indicated changes according to our survey feedback.}
    \label{fig:pattern-dimensions}
\end{figure}

In this article, we describe how we developed the three classes of UX debt mentioned above based on our experience in an industry project.
We arranged them on a spectrum that captures who primarily pays the interest (users vs. developers) and how the debt can be detected (see Figure~\ref{fig:pattern-dimensions}).
We also report the results of company-internal online survey with 21 software practitioners that we conducted to validate our UX debt classes.

With this article, we want to raise awareness in the software engineering research community that not all forms of technical debt are code-centric and that UX debt might even exist without traditional technical debt.
A holistic view on technical debt should also include the human angle, in particular the impact on end users.

\section{Related Work}

Although \emph{code-centric technical debt} has been extensively researched in the past decade~\cite{DBLP:journals/software/CiolkowskiLM21}, broader notions of the concept appear rare~\cite{DBLP:conf/euromicro/StorrleC19, Ahmad2022}.
Since publishing an early draft of this article~\cite{baltes2021ux}, further papers on UX debt appeared~\cite{Rodriguez2023}.
However, in general, technical debt research is very strongly focused on the implementation level.

In 2012, Kruchten et al. mentioned that since Cunningham coined the term, the technical debt metaphor has been used to describe ``many other kinds of debts [...] of software development''~\cite{DBLP:journals/software/KruchtenNO12}.
Although Kruchten et al. mention ``low external quality'', they frame technical debt with a strong focus on code. 
In 2021, Ciolkowski et al. reflected that interest in technical debt ``can affect other internal and external qualities, such as [...] usability'', but that code is ``by far the most studied aspect.''~\cite{DBLP:journals/software/CiolkowskiLM21}.
In 2023, Jaspan and Green noted that the original use of the technical debt metaphor 
``frames technical debt as arising mostly from nontechnical [...] factors'', but they again focus on the code and the ``engineering organizations''~\cite{DBLP:journals/software/JaspanG23b}. 

Fairbanks argues that the modern notion of technical debt diverges from Cunningham's original idea, which was about the benefits and trade-offs of iterative software development~\cite{DBLP:journals/software/Fairbanks20a}.
He proposes the term \emph{ur-technical debt} to refer to the original notion of technical debt.
Störrle and Ciolkowski introduce the term \emph{domain debt} to describe ``the mis-representation of the application domain by an actual system''~\cite{DBLP:conf/euromicro/StorrleC19}. 
Ahmad and Gustavsson remark that debt ``regarding social issues, people, and processes'' has received less attention than more technical aspects~\cite{Ahmad2022}.
They consider this \emph{nontechnical debt} ``inextricably linked'' to technical debt, which aligns with our notion of code-centric UX debt.
In a paper on Android testing, Sun et al. describe an instance of code-centric UX debt where 
a low test coverage of the Android API ``has led to many compatibility issues that can cause apps to crash at runtime on specific Android devices, resulting in poor user experiences''~\cite{DBLP:conf/kbse/SunCZL0022}.
Choma et al. investigated how startups deal with UX feedback, and found that for some startups ``UX issues are usually caught after features launch''~\cite{DBLP:conf/icsob/ChomaMSS0Z22}, 
indicating that in this setting, accumulating unintentional UX debt might be more common than intentional UX debt.
The research closest to ours is that of Rodriguez et al.~\cite{Rodriguez2023}. 
Although they investigated many aspects of UX debt and presented a conceptual model of it, they do not classify UX debt along the same dimensions that we propose (see Figure~\ref{fig:pattern-dimensions}) and instead focus on symptoms and their implications rather than the fact that UX debt can exist without technical debt being present.


\section{Methodology}

Our research involved two main phases: First, based on our own industry experience, we derived three UX debt classes and informally validated them with colleagues working on other projects with different customers. Second, to substantiate and validate our findings, we conducted a detailed company-internal online survey with professional software engineers, with the aim of gauging their perspectives on the UX debt categories we proposed.
The first phase spanned the years 2020 and 2021, while the two authors were working together on a project with a large German telecommunications company. 
The second phase spanned the years 2022 and 2023 and concluded with the online survey in February/March 2023 and a subsequent analysis of the study results.
We provide the questionnaire, quantitative survey data, and scripts used to analyze the data as supplementary material~\cite{SupplementaryMaterial2024}.

\section{Case Study}

We started our research with the assumption that the concepts \emph{technical debt} and \emph{UX debt} are related but different. To gain a better understanding of their relationship, we conducted an exploratory case study while working together on an industrial software project.
We both worked as developers and were responsible for eliciting and documenting requirements, as well as maintaining existing and implementing new features.
In our daily work, we were in close contact with end users through group chats, regular demos, and Q\&A sessions.
Based on that experience, we first defined and then continuously refined three generally applicable classes of UX debt, which we describe in Sections~\ref{sec:code-centric}, \ref{sec:architecture-centric}, and \ref{sec:process-centric}.
We also conducted short informal expert interviews with lead software architects working in the same company but on different projects.
They confirmed that they had encountered instances of the UX debt classes, although to varying degrees. 
This motivated us to conduct an online survey to more systematically collect feedback on the classes (Section~\ref{sec:study}). 

The context of our case study was an industrial (closed source) web application in the domain of data analysis for voice processing systems.
The project had, as of January 2022, almost three years of development time and around 70k lines of code. Over the years, an average of five developers and one to two digital designers were part of our development team.
The front-end was implemented in TypeScript using the Angular UI (user interface) framework, and the back-end was a Spring Boot Java application.
The software had about 150 active expert users.
The overall project our application belonged to was managed using the Scaled Agile Framework (SAFe).

In the following, we introduce each UX debt class along with potential mitigation strategies.
We further report instances of those classes that we experienced in the above-mentioned project.

\subsection{Code-centric UX Debt}
\label{sec:code-centric}


The first class of UX debt, which is closest to the traditional notion of technical debt, captures technical debt that directly affects usability and is usually introduced intentionally. 
We refer to this as \emph{code-centric UX debt}.
In our project, for example, we identified several instances of code duplication that led to inconsistencies in the user-facing behavior of our application. Over time, duplicated CSS classes caused the same UI components to look differently within the application as the copied code fragments diverged (e.g., at some point we had three copied-and-pasted table components that looked slightly different from each other).
This inconsistency in the appearance of functionally equal components led to confusion among users and subsequent requests through our support channels.
Beyond this example, it is important to note that not all CSS clones automatically lead to UX debt.

{\small
\begin{roundedbox}
\noindent\textbf{Class:} \emph{Code-centric UX debt}\\
\textbf{Context:} Development shortcuts that directly correspond to one or more lines/blocks of source code, that is, traditional code-centric technical debt that causes usability issues.\\
\textbf{Problem:} Although the code causing this class of UX debt could be monitored using static analysis tools, the debt that users pay due to the reduced usability of the application is often not factored in.\\
\textbf{Example:} Code clones in CSS classes that cause inconsistent and thus irritating user-facing behavior of front-end components.\\
\textbf{Mitigation:} Configure static analysis tools to scan CSS, HTML, and other UI-focused files (note: functionality is often limited for such files). 
\end{roundedbox}}

In addition to CSS clones, we observed code duplication in the application code that led not only to diverging layouts between functionally similar components but also to diverging functionality. For example, sorting and filtering was possible in one instance of a table component but not in another. This behavior confused users, but also slowed down their workflows because they could not rely on a feature being available in all contexts. 

For this UX debt class, developers pay considerable interest, because at some point the diverging layouts need to be consolidated in a time-intensive refactoring session (which is what we did in our project).
Fortunately, tool support exists for detecting and avoiding UX debt instances of this class because such UX issues are closely tied to related code issues. Exemplary static analysis tools that can detect code clones in CSS classes include Sonar, PMD, and code inspections in modern IDEs such as IntelliJ.
However, their functionality is limited for CSS, HTML, and other UI-focused files.

\subsection{Architecture-centric UX Debt}
\label{sec:architecture-centric}

The second UX debt class, \emph{architecture-centric UX debt}, is still closely related to the implementation level, but the debt is not easily measurable with existing tools~\cite{DBLP:journals/software/CiolkowskiLM21}. This class highlights the trade-off between a ``clean'' software architecture and good usability.

{\small
\begin{roundedbox}
\noindent\textbf{Class:} \emph{Architecture-centric UX debt}\\
\textbf{Context:} Suboptimal composition or placement of front-end components implied by architectural decisions or lack thereof. \\
\textbf{Problem:} Architecture decisions (in the back- or front-end) are made without taking into account how they affect the ability to arrange UI components or their coherence across views.\\
\textbf{Example:} A certain API design or component split is technically more convenient (e.g., eases passing data from UI parent to child component) but it leads to suboptimal placement from a UX perspective.\\
\textbf{Mitigation:} Besides traditional code review, add an additional \emph{domain review} step in which a domain expert evaluates features from a UI/UX perspective, providing early feedback.
\end{roundedbox}}

Modularization and reusable components are essential aspects of a maintainable software architecture.
An example in our project where software architecture and usability competed with each other was the design of certain UI container components.
We designed containers to be as self-contained as possible with a clean event-based interface for inter-component communication.
Those components were, from an architectural perspective, perfectly reusable and well-designed.
However, the chosen architecture made it difficult for us to move contained UI components between the containers because they operated on encapsulated data.
Thus, our clean architecture did not represent a meaningful grouping from the user's point of view.
For example, a developer might place a button in one component because it fits there best from a technical perspective.
However, for a more intuitive user experience, the button should be placed somewhere else, e.g., in a different container.

UX debt in this class usually does not correspond to code-centric technical debt and thus can not be detected using static code analysis alone; user research is required.
Since this class is often reflected in specific views or UI components, we tried to address it with a \emph{domain review} step in our development process, where domain experts provided early feedback during the implementation of a feature, so that architectural decisions and their impact on UX can be critically reflected early in the development process.
Architecture-centric UX debt can be introduced intentionally or unintentionally, depending on whether the developer intentionally breaches common UI patterns or is not aware of users' actual needs. In any way, we assume that users pay the largest part of the interest in this situation.
However, in our empirical study, participants rather considered developers to be the ones paying the largest part. 

\subsection{Process-centric UX Debt}
\label{sec:process-centric}

While for the previous two classes, the UX debt was still closely tied to particular parts of the implementation (but not necessarily detectable using static analysis), the third class captures more abstract UX issues representing a misalignment of how users want to work and the workflows that the software imposes upon them. 
If the development team does not consider an in-depth knowledge of their users' desired processes important, maybe even thinking they knew better, 
this is likely to result in conflicts.
Users are frustrated due to the lack of workflow alignment, and developers are inclined to repeatedly point to the available documentation in case users do not utilize the software as intended.
Although we did not observe such ignorant behavior in our team, a certain degree of misalignment and a know-it-all mentality was indeed present from time to time.
However, it is important to note that, in the original notion of technical debt, there will always be a certain degree of \emph{process-centric UX debt} because one cannot anticipate all potential user workflows; one needs to learn and understand usage over time.

{\small
\begin{roundedbox}
\noindent\textbf{Class:} \emph{Process-centric UX debt}\\
\textbf{Context:} Looking at the application as a whole, it might not support the users' desired and most efficient ways of working.\\
\textbf{Problem:} Whether or not the user-facing parts of an application are aligned with user processes cannot be decided by analyzing the implementation alone. If their workflows are not taken into account during development, the application cannot be used intuitively and efficiently.\\
\textbf{Example:} A common user workflow is spread over several UI components or views of the application.\\
\textbf{Mitigation:} Regularly conduct user research, spanning from early requirements elicitation to validating implemented features using interviews, surveys, or log data analysis.
\end{roundedbox}}

Again, as for the previous class, a software system can be implemented without any code-centric technical debt and yet be completely misaligned when it comes to supporting user processes.
No tool support can fix such issues; the only mitigation is user research and frequent interactions with stakeholders.
Depending on whether developers deliberately decide to ignore user feedback or are simply victims of their cognitive biases, this class represents intentional or unintentional UX debt. The interest is paid almost exclusively by the users and not by the developers.
To detect process-centric UX debt, the entire application must be evaluated.
Teams can use different evaluation methods to catalog existing usability issues.
These methods include observational studies, heuristic evaluations, or survey instruments such as SUS (system usability scale) or NPS (net promoter score).
GitLab, for example, employs SUS as a key performance indicator (KPI)~\cite{GitLab2023}.
Integration of UX research methods into agile process models is sometimes labeled \emph{Lean UX}~\cite{SAFe2023}.


In our project, an example of a mismatch between user workflows and UI layout was the placement and grouping of tabs in our navigation header.
For users, it was not intuitive, as it did not correspond to their usual work process.
It did, however, correspond to our understanding of the workflow.
For users, the layout resulted in cumbersome navigation paths, or even worse, in some cases, they could not find the desired functionality and thought that it was missing, triggering feature requests for features that were already available.
To mitigate such issues, we used telemetry and log data, which is a relatively inexpensive way to understand which features of the software are being used and which are not. 
For example, if a feature is neglected, this can be an indication of UX debt because users might not find the corresponding feature in the UI. 
With such an insight, one can specifically ask about the feature when interacting with users.

In this context, we also want to comment on user documentation and manuals.
Although these documents are supposed to help users become more acquainted with the UI, we experienced two situations where this is not the case: documentation in the wrong place and documentation with the wrong scope.
In our experience, very few users read a manual in advance; instead, they ask for help when a problem arises.
If documentation exists but is not linked directly from the potentially problematic parts of the UI, its value to users is limited.
Even worse is outdated or bloated documentation. If a manual is referenced, but it only confuses the user even more due to the lack or overabundance of information, the frustration is immense. In summary, documentation can sometimes be a cause and not a mitigation of UX debt.




\begin{figure}
    \centering
    \includegraphics[width=0.8\columnwidth,trim=0pt 30pt 10pt 30pt,clip]{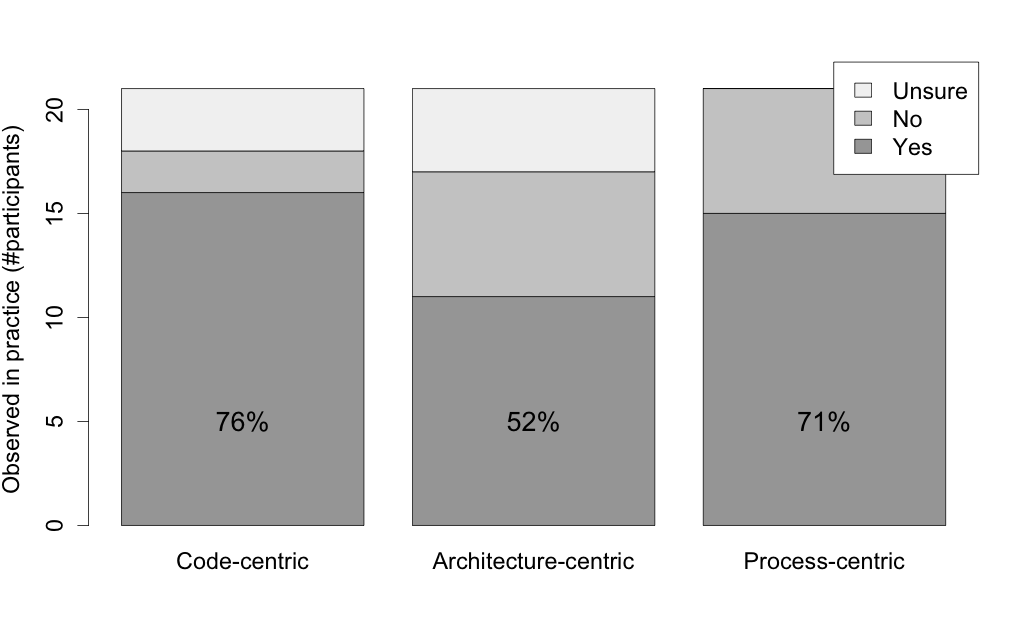}
    \includegraphics[width=0.8\columnwidth,trim=0pt 30pt 10pt 30pt,clip]{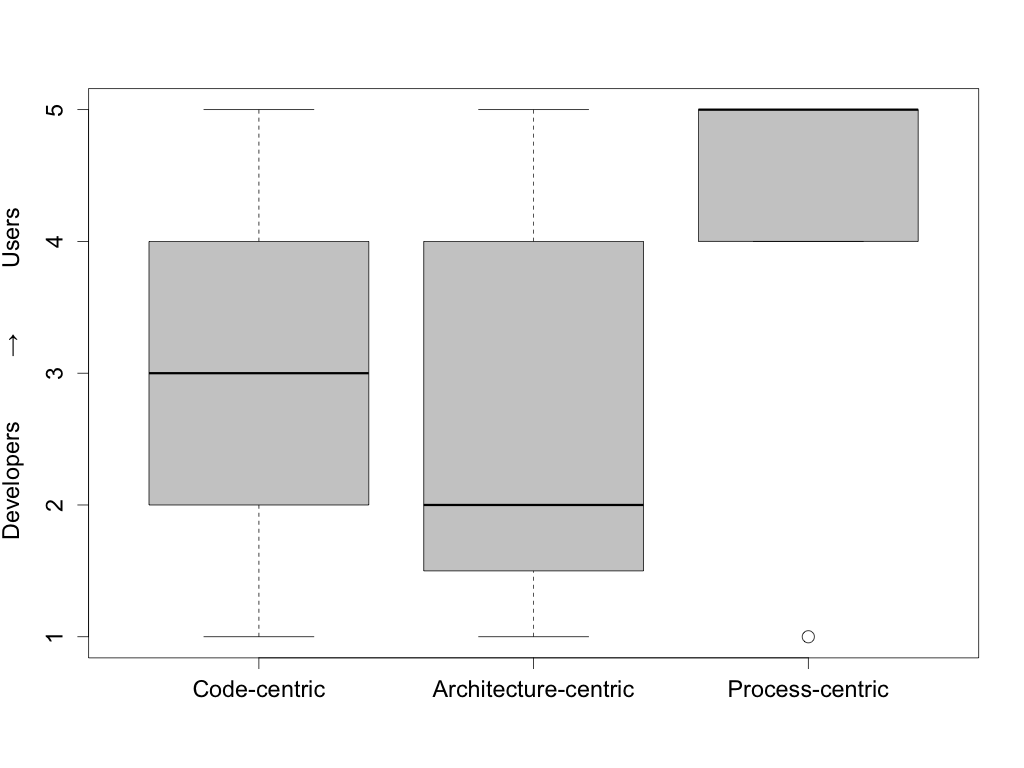}
    \caption{Bar charts show number of participants who observed the corresponding UX debt classes in practice ($n=21$); box plots show participants' ratings for who's primarily paying the debt in the different UX debt classes ($n=21$).}
    \label{fig:survey-results}
\end{figure}

\section{Online Survey}
\label{sec:study}

To validate and refine our UX debt classes, we conducted a survey with software practitioners.
In February 2023, we distributed a call for participation through internal channels at QAware GmbH and were able to recruit 21 people in software development roles (corresponding to a response rate of approximately 10\%).
We asked participants to evaluate whether they had encountered the three classes of UX debt in their daily work.
Participants who observed a class were asked to substantiate their observation with an example.
Those who were uncertain or did not recognize a class were encouraged to explain their ratings.
Finally, participants were asked to rate who primarily pays for a class (users vs. developers).
Of our 21 participants, 16 (76.2\%) worked as developers or architects, three (14.3\%) in requirements engineering roles, one was a development manager, and one had a supporting role.
Their professional development experience ranged from one to more than 30 years. 

Figure \ref{fig:survey-results} shows how many participants observed the corresponding UX debt classes in practice.
Most of the participants (76. 2\%) observed code-centric UX debt; 14.3\% were uncertain and 9.5\% did not face this class of UX debt.
Furthermore, there was a broad agreement for process-centric UX debt, with 71.4\% having observed it in practice.
However, architecture-centric UX debt was more controversial, with only 52.4\% observing it in practice.

To contextualize our quantitative results, we analyzed the open-ended survey answers.
The two authors used a digital whiteboard to group all statements collaboratively.
We first placed all responses as notes on the board and then, in an interactive session, extracted statements, discussed and grouped them.

For \textbf{code-centric UX debt}, a participant clarified his ``unsure'' rating, asking whether a mix of margins and padding in components would qualify as UX debt.
Since we consider this to be an instance of UX debt, this individual has encountered code-centric UX debt, even though they were uncertain about it. 
The participants indicating that they observed this class in practice provided examples that we grouped into the following categories.

{\small
\begin{roundedbox}
\noindent\textbf{Class:} \emph{Code-centric UX debt}\\
\noindent\textbf{Styling:} Participants mentioned problems with styling, especially CSS issues (mentioned 12 times) and inconsistencies (3). In addition, the use of deprecated CSS classes and CSS bleeding was mentioned. 

\noindent\textbf{Code Reuse:} Captures problems such as copy-paste errors (3), 
mismanagement of CSS classes (5), and copying UI components without understanding their usage context or parameters. 

\noindent\textbf{Layout and Design:} Includes various design issues that disrupt UI coherence, e.g., inconsistent indentations (5), layouts (5), or fonts (2).

\noindent\textbf{Tooling/Framework Deficiencies:} Captures challenges related to inadequate CSS frameworks, the absence of tools to maintain CSS quality, and disabled analysis tools, allowing UX debt to remain undetected. 
\end{roundedbox}}

Open-ended responses to \textbf{architecture-centric UX debt} varied widely among survey participants, indicating a lack of consensus or understanding of this class. 
We grouped the responses of the users who observed the class into two categories, where the second category can be considered a subcategories of the first.

{\small
\begin{roundedbox}
\noindent\textbf{Class:} \emph{Architecture-centric UX debt}\\
\noindent\textbf{Back-end/Front-end Misalignment:} This includes issues arising from a back-end-centric design approach, which might separate data that users expect to see together, complicating back-end requests for related UI components. Participants also mentioned poor performance caused by a large number of backend requests. This category captures seven statements similar to this one: ``Components were mainly designed [...] starting with the backend data and not the user workflows.''

\noindent\textbf{Conway's Law:} Back-end APIs and front-end components might be arranged along the responsible teams and not based on user needs, resembling Conway's law~\cite{Conway1968}.
Although the quantitative agreement with this UX debt class was the lowest, we received many examples for the above-mentioned back-end/front-end misalignment, providing support that this class does indeed exist and is relevant in practice.
\end{roundedbox}}

Unfortunately, we only received four comments from participants who did not observe the class or were uncertain, two of whom mentioning that they simply had not observed the class yet.
One pointed out that inconsistent UI components can have many causes and are not necessarily the result of architecture decisions, which is, of course, correct. 
However, this does not exclude the possibility that UX debt can be introduced by purely technical architecture decisions.
Another participant outlined that a ``good'' software architecture should be closely coupled with the user workflow so that this kind of UX debt does not occur.
Although this is true, we would argue that this is not always the case in reality.

For \textbf{process-centric UX debt}, two participants who had not observed the class provided an open-ended response, with one participant indicating that they only experienced it from a user perspective. 
From the responses of participants who observed process-centric UX debt in practice, we derived two categories. 

{\small
\begin{roundedbox}
\noindent\textbf{Class:} \emph{Process-centric UX debt}\\
\noindent\textbf{Broken Flows:} Captures eight answers reporting unnecessary steps caused by the UI not reflecting user workflows, e.g., copying-and-pasting data between forms, required re-authentication within the same application, or missing keyboard shortcuts. 
One participant wrote: ``[The user] has to do several clicks to get the job done.'' 

\noindent\textbf{Search and Navigation:} Users face UX issues due to limited search capabilities or lack of fast attribute-based filtering. 
\end{roundedbox}}


Figure \ref{fig:survey-results} shows the ratings of our participants on who is primarily paying for each UX debt class.
Compared to our classification, one can see that participants considered architecture-centric UX debt to have a stronger developer focus.
Process-centric UX debt is clearly categorized as being primarily paid by users and hence aligned with our initial classification. 
The open-ended responses provide an explanation for diverging interpretations of architecture-centric UX debt. 
Respondents assumed that a good architecture is so flexible that it can always be adapted to the users' processes and that if this is not the case, the software architecture is broken and, therefore, the developers suffer more than the users.
However, we argue the other way around: developers put a technically refined architecture decision above the processes of the users, and therefore the solution is optimal from an architectural point of view, but not necessarily for the user.
Since all of our participants were from the same company, a broader study is needed to shed more light on this UX debt class.

\section{Conclusion}

Based on our industry experience, we derived three classes of UX debt: \emph{code-centric}, \emph{architecture-centric}, and \emph{process-centric UX debt}. We aligned these classes on two dimensions: the primary payer of the debt (users vs. developers) and suitable means for debt detection (code analysis vs. user research). 
In addition, we conducted a survey with practitioners that confirmed the relevance of these three classes and provided additional context.
Currently, most research still targets code-centric technical debt, focusing on the developer perspective.
We argue for a broader perspective that takes end-users into account, especially because they are often the ones (also) paying the interest of intentionally or unintentionally introduced UX debt. 
Similar to code-centric technical debt, the more UX debt is built up, the harder it is to remove, which for UX debt means making the software usable again. 
Until UX debt is removed, considerable resources are wasted because users work inefficiently and get frustrated.
This might even lead to them losing trust in the software and, as a consequence, to a high churn rate if users switch to competitors.
We want to encourage the software engineering research community to both empirically study UX debt in detail and to work on novel tool support to detect and quantify UX debt early in the development process.
One potential avenue could be to develop visualizations that display UX debt measured using log analysis, usability metrics, and other forms of user research, along with traditional metrics that capture code-level technical debt.



\clearpage



\bibliographystyle{ACM-Reference-Format}
\bibliography{literature}


\end{document}